\documentclass{ws-procs961x669}            
\begin{document}
\title{A simple parametrisation for coupled dark energy}

\author{Vitor da Fonseca$^*$ and Nelson J. Nunes}

\address{Instituto de Astrofísica e Ciências do Espaço, \\Faculdade de Ciências da Universidade de Lisboa,\\
Campo Grande, PT1749-016 Lisboa, Portugal\\
$^*$E-mail: fc52156@alunos.fc.ul.pt}

\author{Tiago Barreiro}

\address{Instituto de Astrofísica e Ciências do Espaço,\\ECEO, Universidade Lusófona de Humanidades e Tecnologias,\\
Campo Grande, 376, 1749-024 Lisboa, Portugal\\}

\begin{abstract}
We propose a phenomenological generalisation of the standard model with only one extra degree of freedom that parametrises the evolution of a scalar field responsible for the cosmic acceleration. The model also foresees an additional parameter in the form of a coupling between dark energy and dark matter. This model captures a large diversity of dark energy evolutions at low redshift  and could usefully complement common CPL parametrisations widely used. In this context, we have been constraining the parametrisation with data from Planck and KiDS, bringing different results between the early and late universe observations.
\end{abstract}

\keywords{Quintessence, dark matter, observational cosmology.}

\bodymatter

\section{Introduction}
\label{sec:intro}

The discovery in 1998 that the Universe is accelerating  \cite{acel1,acel2} was really stunning at the time. While the source of the cosmic acceleration is yet to be unveiled, the existence of a dark energy component has been postulated to explain it \cite{amendola,mod}. In the standard model, dark energy takes the form of a cosmological constant $\Lambda$ which by definition possesses a constant energy density and negative pressure that fights the attractive pull of matter \cite{cc1}. That model is, at the moment, compatible with all observations and it is the simplest one \cite{Planck2018}, at least from a parametrisation point of view. One of the outstanding issue in the concordance model is the so-called coincidence problem \cite{martin}. Given how the energy densities dilute with the expansion, the fact that the cosmological constant is catching up with matter at the present time necessitates initial conditions at the origin that are dramatically precise. This issue has been paving the way to study varying dark energy \cite{scalar3} in order to reproduce more dynamically the current abundances.

The simplest and model independent approach is phenomenological. It consists in parametrising the dark energy equation of state, being the pressure over the energy density that equals $-1$ for a cosmological constant. The detection of a time dependence would exclude a cosmological constant.  Introducing the fewest possible parameters is ideal to limit degeneracies between them. A simple and popular parametrisation \cite{param_z} is a Taylor expansion at first order which has a very limited domain of validity. Various improvements have been introduced, like the Chevallier-Polarski-Linder (CPL) parametrisation \cite{param_a1,param_a2}. Also, to account for the possibility of early dark energy, a parametrisation stepwise in redshift \cite{Corasaniti} has been proposed with a larger number of additional parameters.

Here, we explore an alternative way to parametrise dark energy by proposing to parametrise a scalar field $\phi$ called quintessence that would be responsible for dark energy both in the early and late Universe \cite{Wetterich:1994bg}. Instead of setting  an equation of state parameter evolution, or a scalar potential, this method parametrises the dark energy scalar field itself. We argue that with a single parameter we recover the generic behaviour of the previous parametrisations. We test this approach with data from Planck and KiDS to constrain its specific parameters, identifying contrasting results between the early and late universe observations. This is a comprehensive work going from the analytical solutions to parameter constraints.

\section{Scalar field parametrisation}
\label{sec:parametrisation}

We consider that the cosmological fluid is composed of matter (m), i.e. baryons (b) and cold dark matter (c), as well as the scalar field itself in a flat Friedmann-Lemaitre-Robertson-Walker background. The scalar field is supposed to be canonical and  homogeneous at large scale \cite{Caldwell_1998,tsuq}, approximated by a perfect fluid. We choose to work with the number of e-folds, $N\equiv\ln a$, as the time variable. In the following, the prime stands for the derivative with respect to the number of e-foldings, and the dot the derivative with respect to cosmic time.

By combining the Friedmann equation,
\begin{equation}
H^2=\frac{\kappa^2}{3}\left(\rho_b+\rho_c+\rho_\phi\right),
\label{eq:friedmann}
\end{equation}
with the continuity equation of the scalar field,
\begin{equation}
\rho_\phi^\prime+3H^2\phi^{\prime 2}=0, \label{eq:phi}
\end{equation}
and the continuity equation of matter,
\begin{equation}
\rho_c^\prime+3\rho_c=0, \label{eq:cdm}
\end{equation}
one obtains \cite{Nunes:2003ff} a differential equation for the dark energy density (where we conventionnally set $\kappa=1$),
\begin{equation}
\frac{\rho_\phi^\prime}{\rho_{m_0}}=-\phi^{\prime^2}\left(a^{-3}+\frac{\rho_\phi}{\rho_{m_0}}\right),
\label{eq:diff}
\end{equation}
whose general solution is our starting point,
\begin{equation}
\frac{\rho_\phi}{\rho_{m_0}}=e^{-\int_0^NdN\phi^{\prime^2}}\times\left[\frac{\Omega_{\phi_0}}{\Omega_{m_0}}-\int_0^NdN\phi^{\prime^2}e^{-3N+\int_0^NdN\phi^{\prime^2}}\right].
\label{eq:diff_sol}
\end{equation}
Among possible parametrisations enabling us to solve the former equation analytically,
\begin{align}
\phi^\prime=&\epsilon\phi,\\
\phi^\prime=&\sigma\phi^{-1},\\ 	
\phi^\prime=&\lambda,
\label{eq:sol}
\end{align}
where $\epsilon$, $\sigma$ and $\lambda$ are given constants, we explore the simplest one in Eq.~(\ref{eq:sol}) \cite{Nunes:2003ff}.

The advantage of this parametrisation is three folds. Firstly, it is very simple, as we only add one single parameter to extend the standard model. It allows to reconstruct analytically the form of the potential \cite{Nunes:2003ff},
\begin{equation}
V(\phi)=A\,e^{-\frac{3}{\lambda}\phi}+B\,e^{-\lambda\phi},
\label{eq:potential}
\end{equation}
with the following mass scales,
\begin{align}
A &=\frac{3}{2}\,\frac{\lambda^2}{3-\lambda^2}H_0^2\,\Omega_{m_0}\,e^{\frac{3}{\lambda}\phi_0},\\
B &=\frac{3}{2}\,\frac{\lambda^2-6}{3-\lambda^2}H_0^2\left(\frac{\lambda^2}{3}-\Omega_{\phi_0}\right)e^{\lambda\phi_0},
\label{eq:mass_scales}
\end{align}
where  $H_0$ is today's Hubble expansion rate and $\Omega_{i_0}=\rho_{i_0}/3H_0^2$ today's abundance parameter of species $i$. Having a simple form of the potential allows a straightforward implementation of the model within the existing Boltzmann code CLASS\cite{Class1,Class2} that can be validated by comparing the numerical results with the expected asymptotic behaviour of the analytic expressions. Given the small number of parameters involved and thanks to its simplicity, our analysis permits to effectively constrain the model parameters with current data as in the next section.

\begin{figure}[h!]
\centering
  \includegraphics[width=0.5\linewidth]{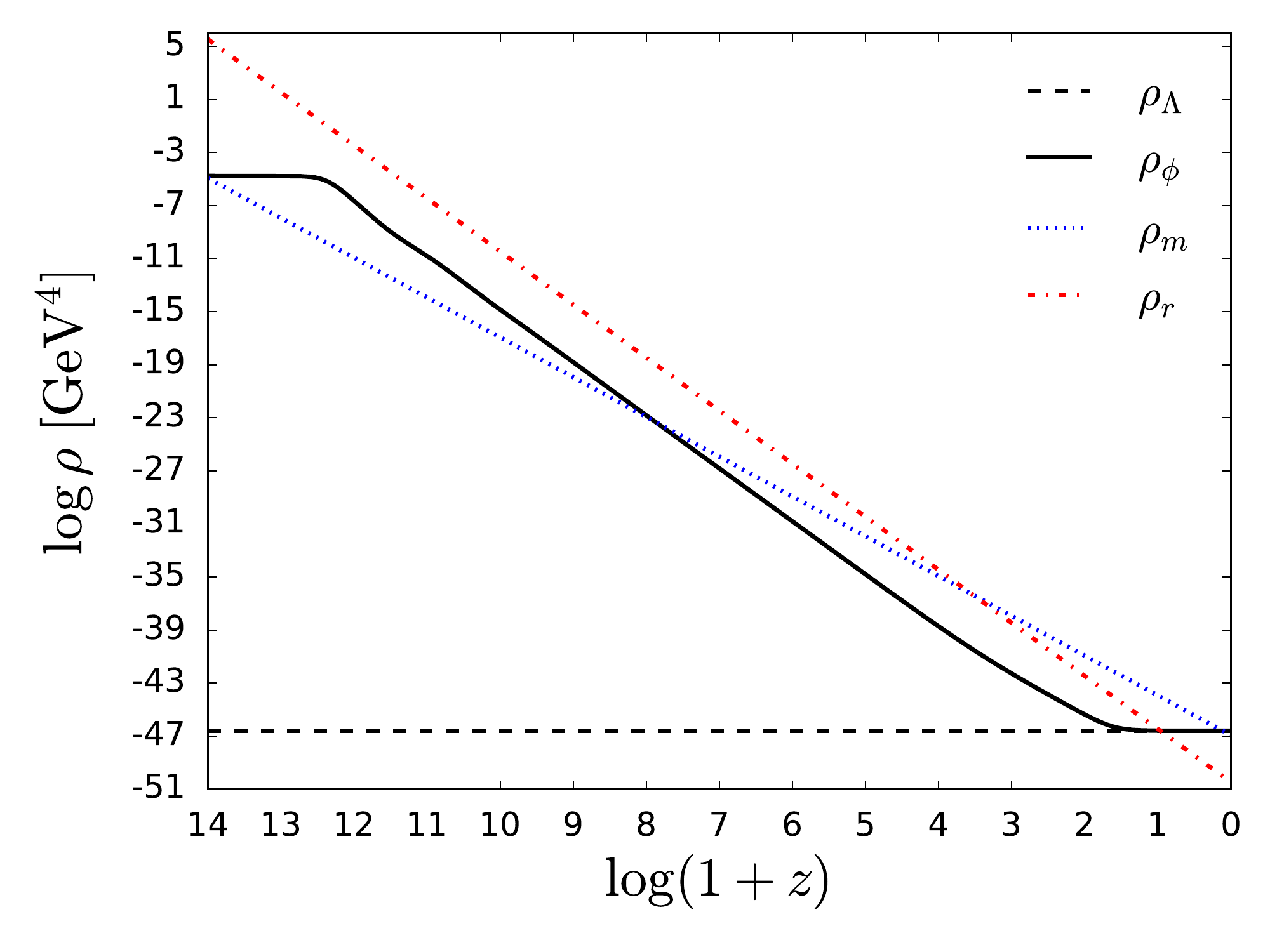}
  \caption{Energy density dilution of a cosmological constant, as well as quintessence, matter and radiation for the scalar field parametrisation ($\lambda=0.01$).}
 \label{fig:scaling}
\end{figure}

Secondly, this scalar potential, being the sum of two exponential terms, does alleviate the severe fine-tuning of the initial conditions as illustrated in Fig.~\ref{fig:scaling}.  For small values of the parameter $\lambda$, when the first exponential term in the potential is steep enough, the scalar field energy density is attracted and temporarily scales with the background as the universe expands, be it dominated by radiation and then matter. Later, thanks to the second shallow exponential, whose slope is the inverse of the first one, the scalar field energy density eventually freezes, mimicking a cosmological constant. By freezing, dark energy overcomes matter at late time, enabling the acceleration of the universe that we are currently experiencing. This well-known attractor mechanism \cite{Barreiro:1999zs} ensures that one can start from a large domain of initial conditions for the scalar field, to reach unavoidably the tracking solution and end up with the correct order of magnitude both during matter and dark energy domination.

\begin{figure}
\centering
  \includegraphics[width=0.5\linewidth]{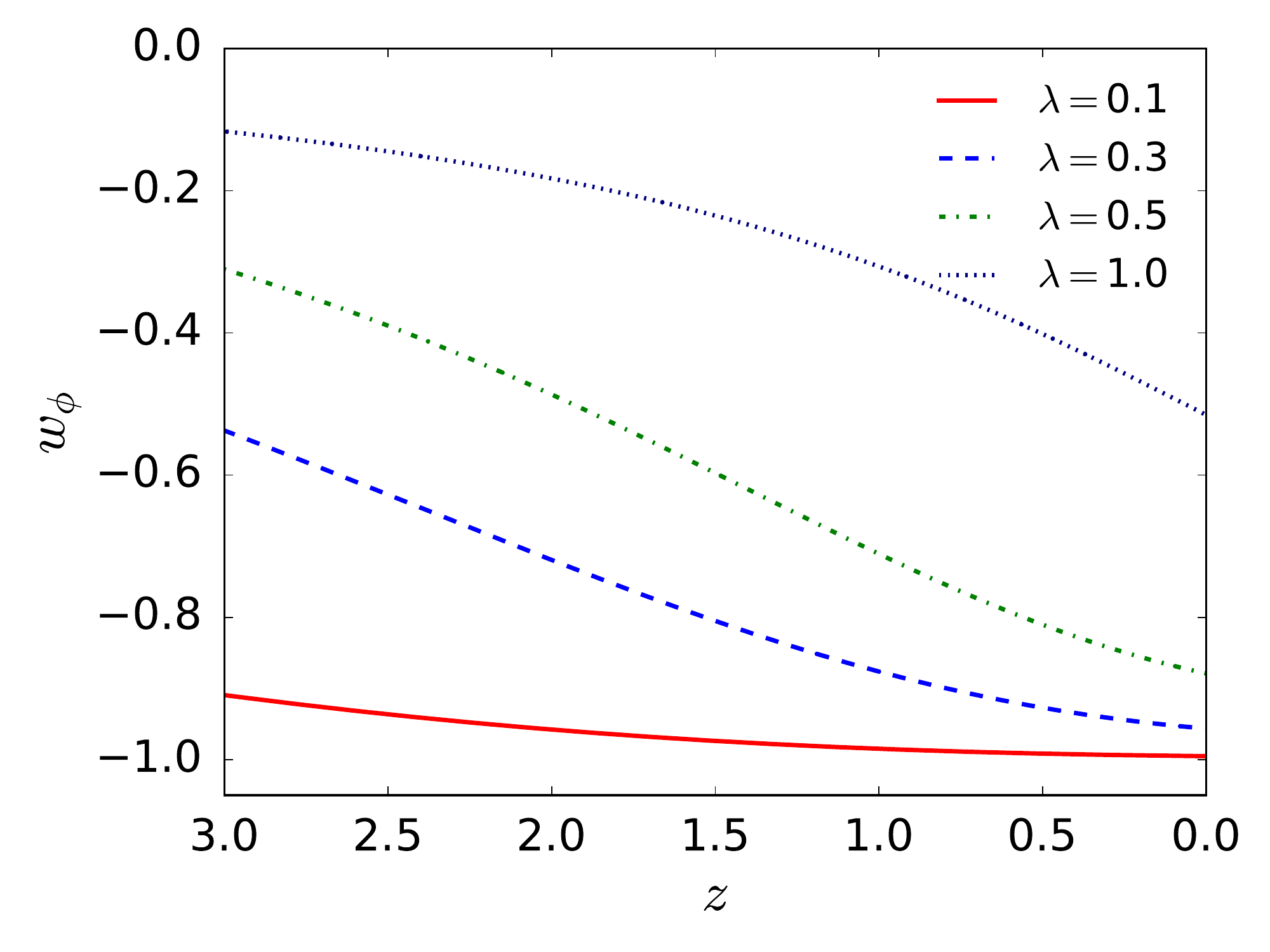}
   \caption{Dynamics of the dark energy equation of state for given values of the parameter $\lambda$.}
   \label{fig:diversity}
\end{figure}

Last but not least, despite just one extra parameter, the model covers a wide range of possible dark energy evolution at low redshift as displayed in Fig~\ref{fig:diversity}. The parametrisation is thus able to capture a large variety of possible dynamics for dark energy. The observational data might even become precise enough in the future for discriminating between these different evolutions.

As a step further, we decide to complement our parametrisation by assuming a non-minimally coupling between dark energy and cold dark matter \cite{PhysRevD.62.043511}. Again for the sake of simplicity, we consider a conformal and constant coupling $\beta$ that parametrises the interaction between dark matter and the scalar field. The stress-energy tensor is jointly conserved within the dark sector in order to ensure the covariance of the theory,
\begin{align}
\nabla_\mu\left(T^{(\phi)}\,^{\mu}_{\nu}+T^{(c)}\,^{\mu}_{\nu}\right)&=0, \\
\nabla_\mu T^{(\phi)}\,^{\mu}_{\nu}&=+\beta\rho_c\nabla_\mu\phi, \\
\nabla_\mu T^{(c)}\,^{\mu}_{\nu}&=-\beta\rho_c\nabla_\mu\phi.
\end{align}
Assuming that dark energy does not interact with baryons, the latter are separately conserved contrary to dark matter and quintessence,
\begin{align}
\label{eq:baryons}
\rho_b&=\rho_{b_0}~e^{-3N}, \\
\label{eq:cdm_coupled}
\rho_c&=\rho_{c_0}~e^{-3N+\beta\left(\phi-\phi_0\right)}, \\
 \rho_\phi^\prime+3H^2\phi^{\prime 2} &=-\beta\phi^\prime\rho_c.
 \label{eq:phi_coupled}
\end{align}
The sign of the product $\beta\phi^\prime$, i.e. $\beta\lambda$ in our parametrisation, determines whether energy is being pumped from the scalar field or from dark matter. Similarly to Ref.~\citenum{Nunes:2003ff}, we use the Friedmann equation (\ref{eq:friedmann}) with the solution we find for the scalar field energy density,
\begin{align}
\rho_\phi=&\left(\rho_{\phi_0}+\frac{\lambda^2+\beta\lambda}{\lambda^2+\beta\lambda-3}~\rho_{c_0}+\frac{\lambda^2}{\lambda^2-3}~\rho_{b_0}\right)e^{-\lambda^2N} \nonumber\\
&-\frac{\lambda^2+\beta\lambda}{\lambda^2+\beta\lambda-3}~\rho_{c_0}\,e^{(\beta\lambda-3)N}
-\frac{\lambda^2}{\lambda^2-3}~\rho_{b_0}\,e^{-3N},
\end{align}
to reconstruct, for the first time, the corresponding potential with similar properties of the dynamical system as in the minimally coupled case,
\begin{equation}
V(\phi)=A\,e^{\left(-\frac{3}{\lambda}+\beta\right)\phi}+B\,e^{-\lambda\phi}+C\,e^{-\frac{3}{\lambda}\phi},
\label{eq:potential_coupled}
\end{equation}
where the mass scales are now the following,
\begin{align}
A &=\frac{3}{2}\,\frac{\lambda^2+2\beta\lambda}{3-\lambda^2-\beta
\lambda}~H_0^2\,\Omega_{c_0}\,e^{\left(\frac{3}{\lambda}-\beta\right)\phi_0},
\label{eq:A_beta}\\
B &=\frac{6-\lambda^2}{2}\,H_0^2\left(1+\frac{3}{\lambda^2-3+\beta\lambda}\Omega_{c_0}+\frac{3}{\lambda^2-3}\Omega_{b_0}\right)e^{\lambda\phi_0},
\label{eq:B_beta}\\
C &=\frac{3}{2}\,\frac{\lambda^2}{3-\lambda^2}~H_0^2\,\Omega_{b_0}\,e^{\frac{3}{\lambda}\phi_0}.
\label{eq:C_beta}
\end{align}

\begin{figure}
\centering
{\includegraphics[width=0.5\linewidth,keepaspectratio]{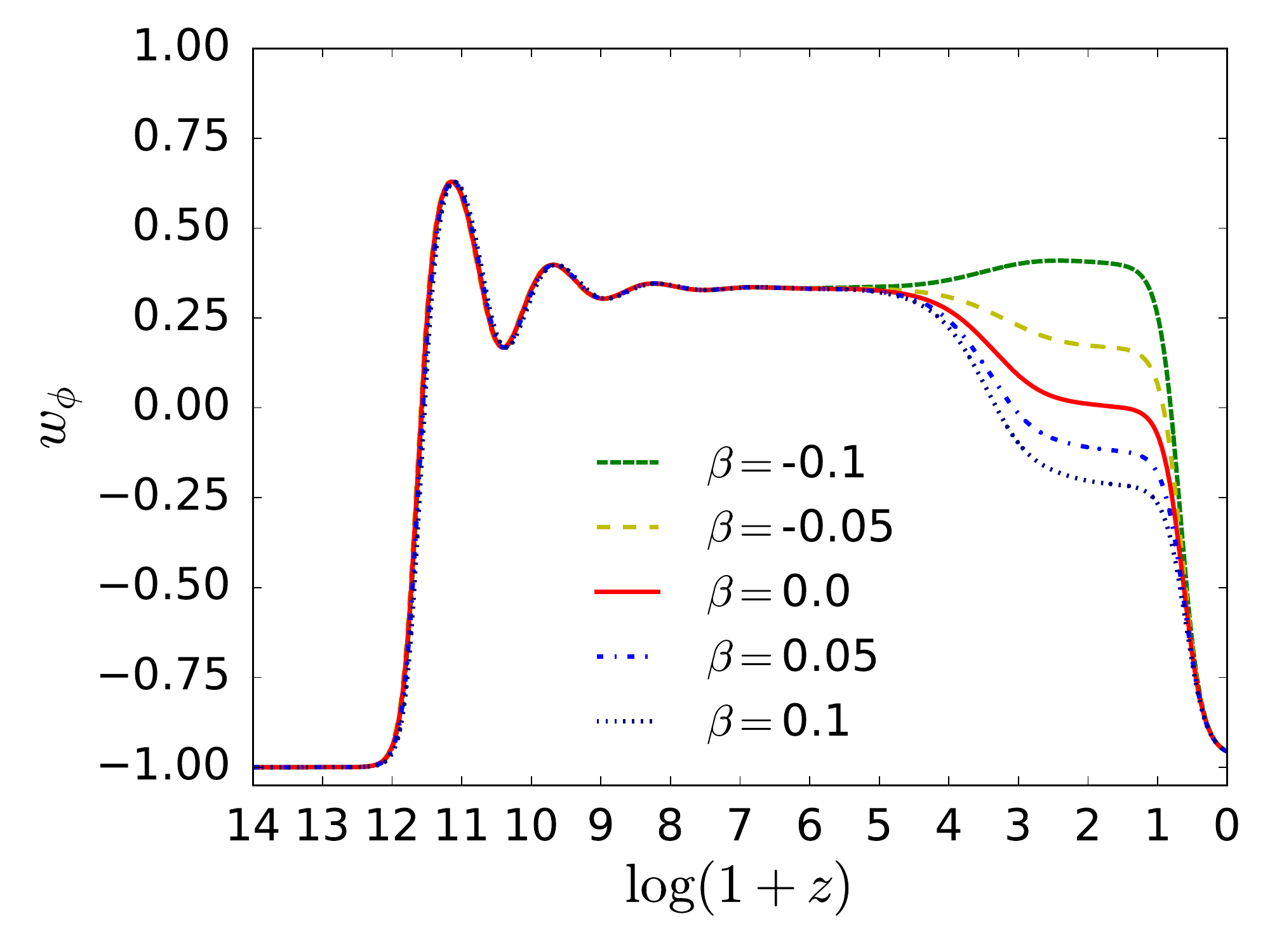}} 
\caption{Dark energy evolution in the coupled parametrisation. Beta negative corresponds to the transfer of energy from the dark matter component to the scalar field.}
\label{fig:eos_beta}
\end{figure}

We modify the CLASS code to accommodate that potential as well as the coupled equations (\ref{eq:cdm_coupled}) and (\ref{eq:phi_coupled}) to simulate the model with the two additional parameters $\lambda$ and $\beta$. The numerical results plotted in Fig.~\ref{fig:eos_beta} are in line with the asymptotic solution for the dark energy equation of state during matter domination when baryonic matter is neglected,
\begin{equation}
w_\phi\approx-\frac{\beta}{\lambda+\beta}.
\label{eq:scaling_matter}
\end{equation}
One can note that today's equation of state, $w_0$, does not depend on the coupling, and drops close to $-1$ during the scalar field dominated epoch for small values of $\lambda$,
\begin{equation}
w_{_0}=-1+\frac{\lambda^2}{3\Omega_{\phi_0}}.
\end{equation}

Furthermore, in order to numerically predict observables that we use to constrain the parameters with observational datasets at the perturbative level, we again modify CLASS to implement the coupled evolution of the scalar fluctuations at linear level in the synchronous gauge adopting usual conventions \cite{Ma_1995}. We modify the existing perturbed Klein-Gordon equation, as well as the equation of motion for the dark matter density contrast $\delta_c$ and velocity perturbations $\theta_c$ to account for the coupling, by adding the relevant sources term in the right-hand side,
\begin{align}
\ddot{\varphi}+2\mathcal{H}\dot{\varphi}+\left(k^2+a^2V_{,\phi\phi}\right)\varphi+\frac{\dot{h}\dot{\phi}}{2}&=-\beta a^2\bar{\rho}_c\delta_c,\\
\dot{\delta}_c+\theta_c+\frac{\dot{h}}{2}&=\beta\dot{\varphi},\\
\dot{\theta}_c+\mathcal{H}\theta_c&=\beta\left(k^2\varphi-\dot{\phi}\theta_c\right),
\end{align}
where the overbar denotes the background quantities. Additionally, we adapt the transformation from the synchronous gauge into the comoving gauge to maintain the matter density source function $\delta_m$ gauge-invariant,
\begin{equation}
\delta_m^C=\frac{\delta\rho_m}{\bar{\rho}_m}+\left(3-\frac{\beta\phi^\prime\bar{\rho}_c}{\bar{\rho}_c+\bar{\rho}_b}\right)\frac{\mathcal{H}}{k^2}\theta_m.
\end{equation}

In parallel, to provide for the code validation, we derive the analytic equation of motion of the dark matter fluctuations in the Newtonian limit during the domination of matter, precisely when the coupling makes a difference,
\begin{equation}
\delta^{\prime\prime}_c+\frac{1}{2}\left(1+3\lambda\beta\right)\delta^\prime_c-\frac{3}{2}\left(1-\frac{\lambda^2}{3}-\frac{\lambda\beta}{3}\right)\left(1+2\beta^2\right)\delta_c=0.
\end{equation}
In the left panel of Fig~\ref{fig:f2}, we note that the scalar field slows the growth of perturbations against the standard model, by decreasing the dynamical term in the equation of motion. On the other hand, the existence of the coupling brings competitive effects that further slow the fluctuation or, on the contrary, accelerate them, depending on whether energy is being transferred from the scalar field or into it. We are also able to get an analytic expression of the growth rate $m+$ under the Newtonian approximation,
\begin{equation}
m_+=\frac{1}{4}\left[-1-3\lambda\beta+\sqrt{24\left(1-\frac{\lambda^2}{3}-\frac{\lambda\beta}{3}\right)\left(1+2\beta^2\right)+\left(1+3\lambda\beta\right)^2}~\right].
\end{equation}
One can see in the right panel of Fig.~\ref{fig:f2} that the numerical results in the matter era on a small scale are very much in line with it. The standard growth, which equals $1$, is obviously recovered for vanishing values of the parameters $\lambda$ and $\beta$.
\def\figsubcap#1{\par\noindent\centering\footnotesize(#1)}
\begin{figure}[h]%
\begin{center}
  \parbox{2.4in}{\includegraphics[width=2.4in]{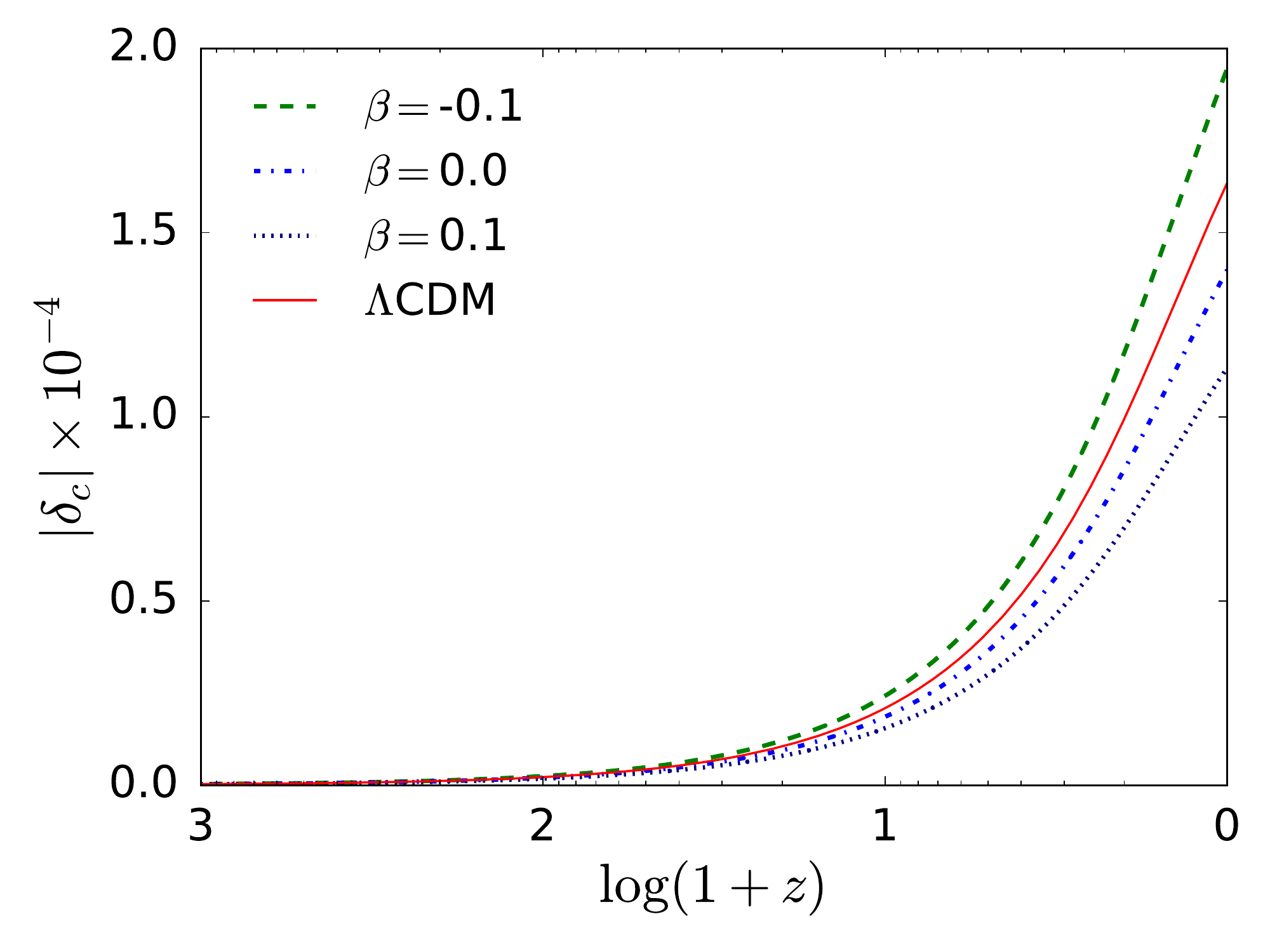}\figsubcap{a}}
  \hspace*{4pt}
  \parbox{2.4in}{\includegraphics[width=2.4in]{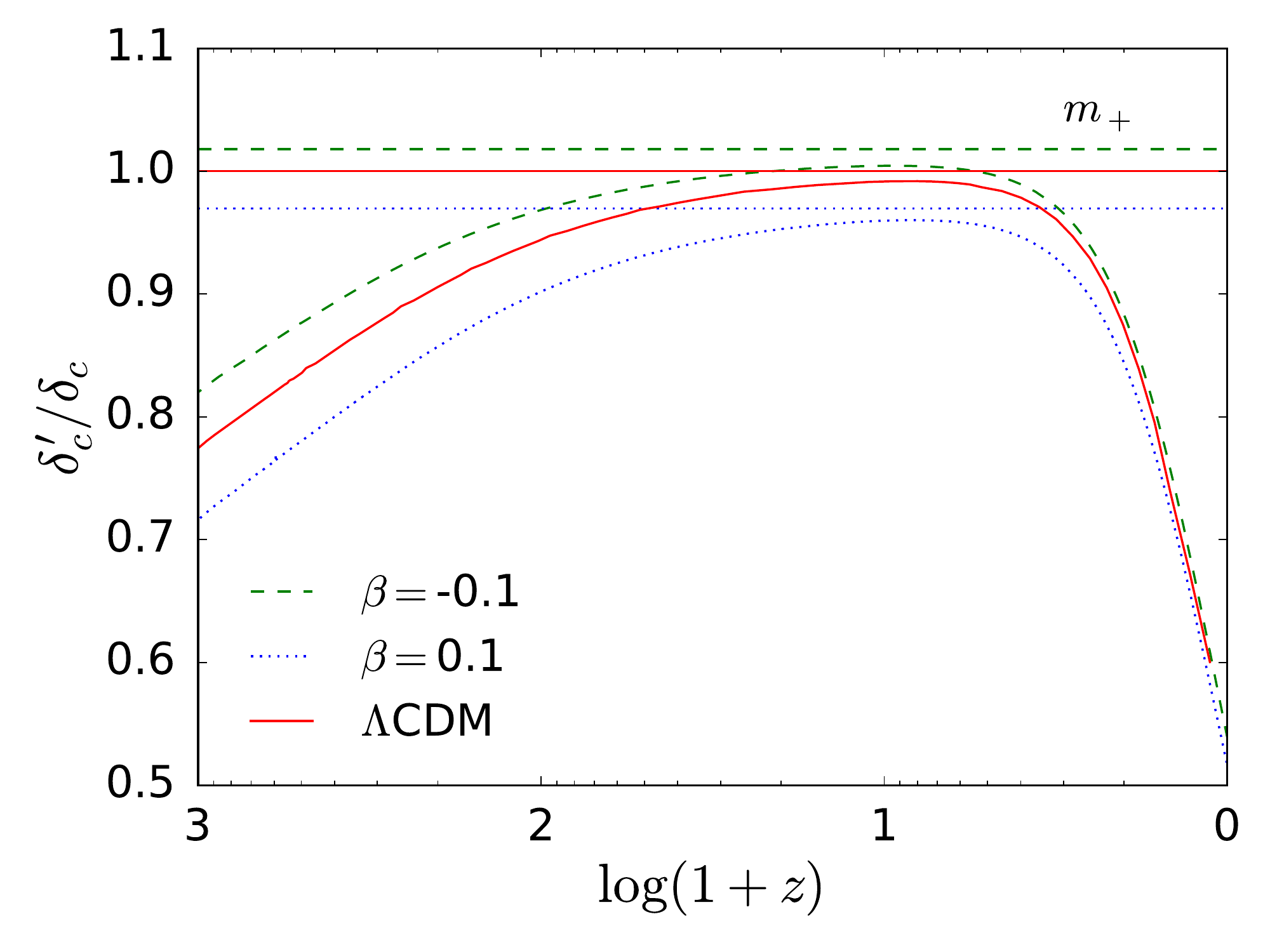}\figsubcap{b}}
  \caption{$\lambda=0.3$ on scale $k=0.1~h$/Mpc. (a) Dark matter fluctuations. (b) Growth rate function.}%
  \label{fig:f2}
\end{center}
\end{figure}
\section{\label{sec:constraints}Observational constraints}

We turn now our attention to the observations in order to constrain the value of the posteriors $\lambda$ and $\beta$ through Bayesian inference with MontePython \cite{MP1,MP2,Feroz_2008,Feroz_2009,Feroz_2019,Buchner:2014nha} and GetDist \cite{Lewis:2019xzd}. Given that it affects the growth of dark matter perturbations, our parametrisation leaves observational signatures, both on the CMB anisotropies and matter power spectrum that we can numerically predict with our adapted version of CLASS and compare with observations.

As for the CMB anisotropies, early dark energy as well as the coupling affect the amplitude and position of the acoustic peaks as illustrated on the left panel of Fig.~\ref{fig:power_spectra}. We use the data from the Planck 2018 likelihood \cite{collaboration2019planck} on the temperature and polarisation of the CMB in the MCMC analysis. As regards today's linear matter power spectrum, the delay or the acceleration in the growth of dark matter perturbations has consequences too (right panel of Fig.~\ref{fig:power_spectra}). We use the gravitational weak lensing measurements, cosmic shear, from the KiDS-450 survey \cite{Hildebrandt_2016} in the Bayesian fitting.
\def\figsubcap#1{\par\noindent\centering\footnotesize(#1)}
\begin{figure}[h]%
\begin{center}
  \parbox{2.4in}{\includegraphics[width=2.4in]{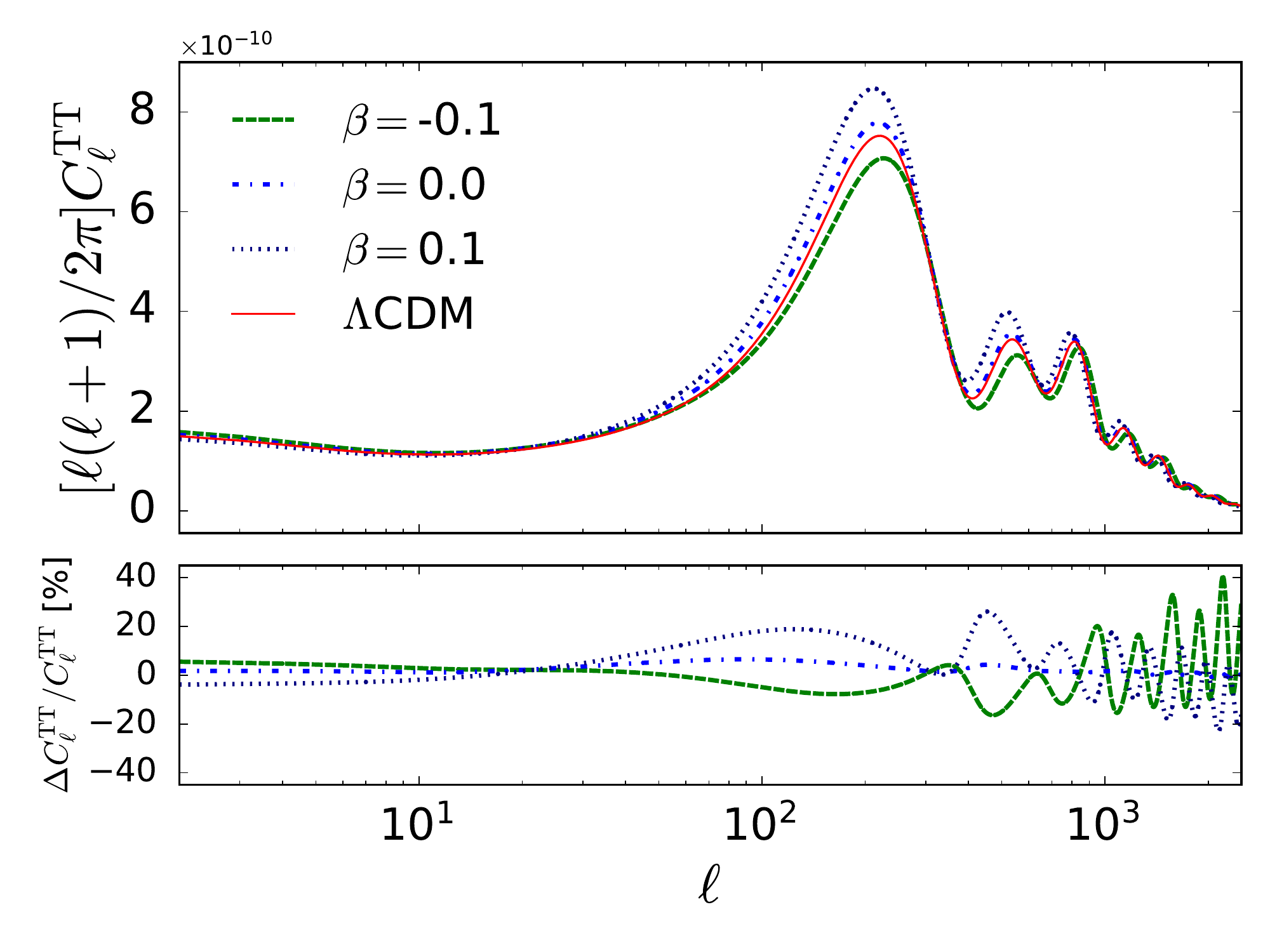}\figsubcap{a}}
  \hspace*{4pt}
  \parbox{2.4in}{\includegraphics[width=2.4in]{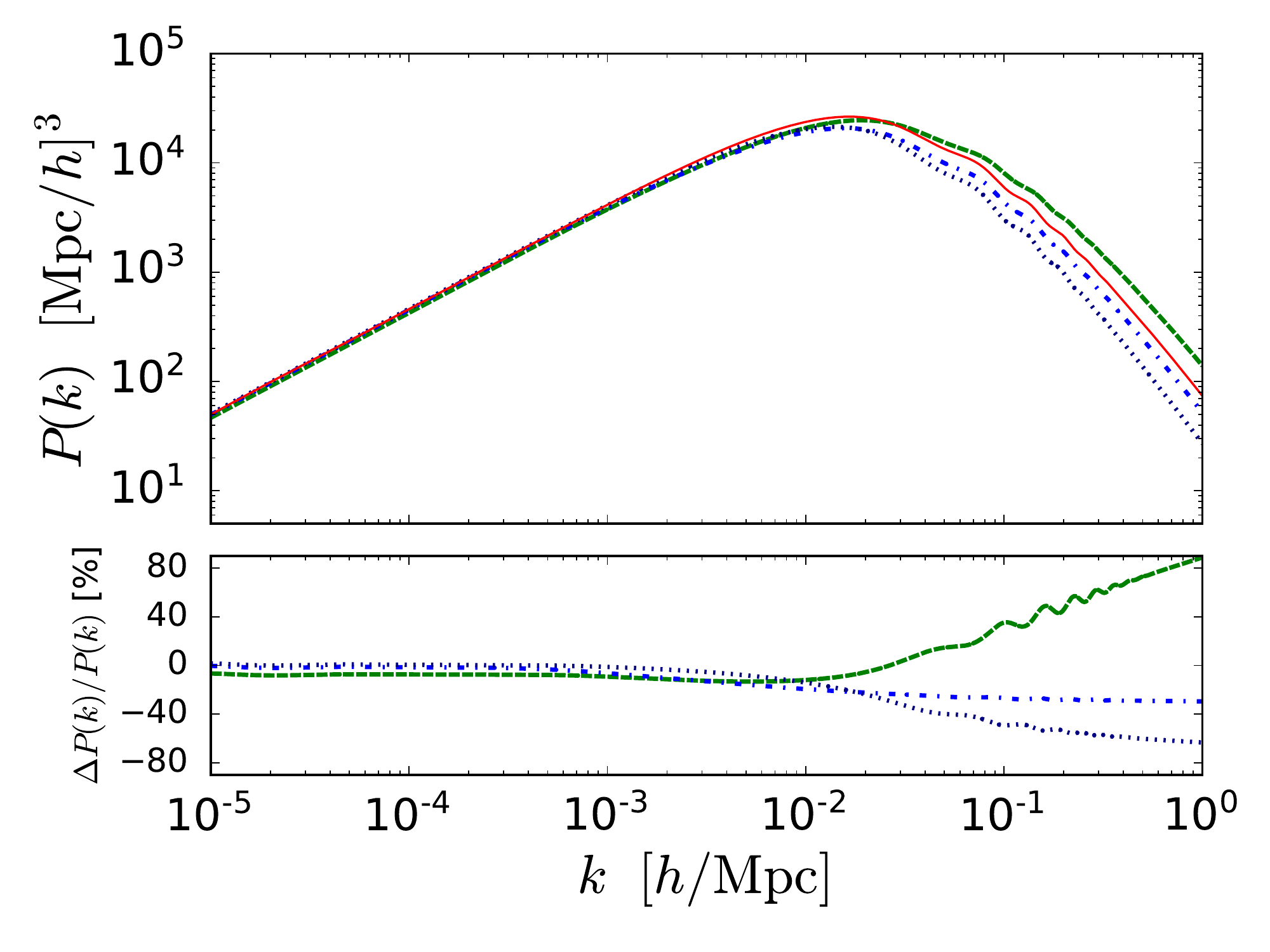}\figsubcap{b}}
  \caption{Observables predictions for $\lambda=0.3$ (a) CMB dimensionless angular power spectrum.  (b) Linear matter power spectrum at redshift $z=0$.}%
  \label{fig:power_spectra}
\end{center}
\end{figure}

\begin{figure}[h]
\begin{center}
\includegraphics[width=2.5in]{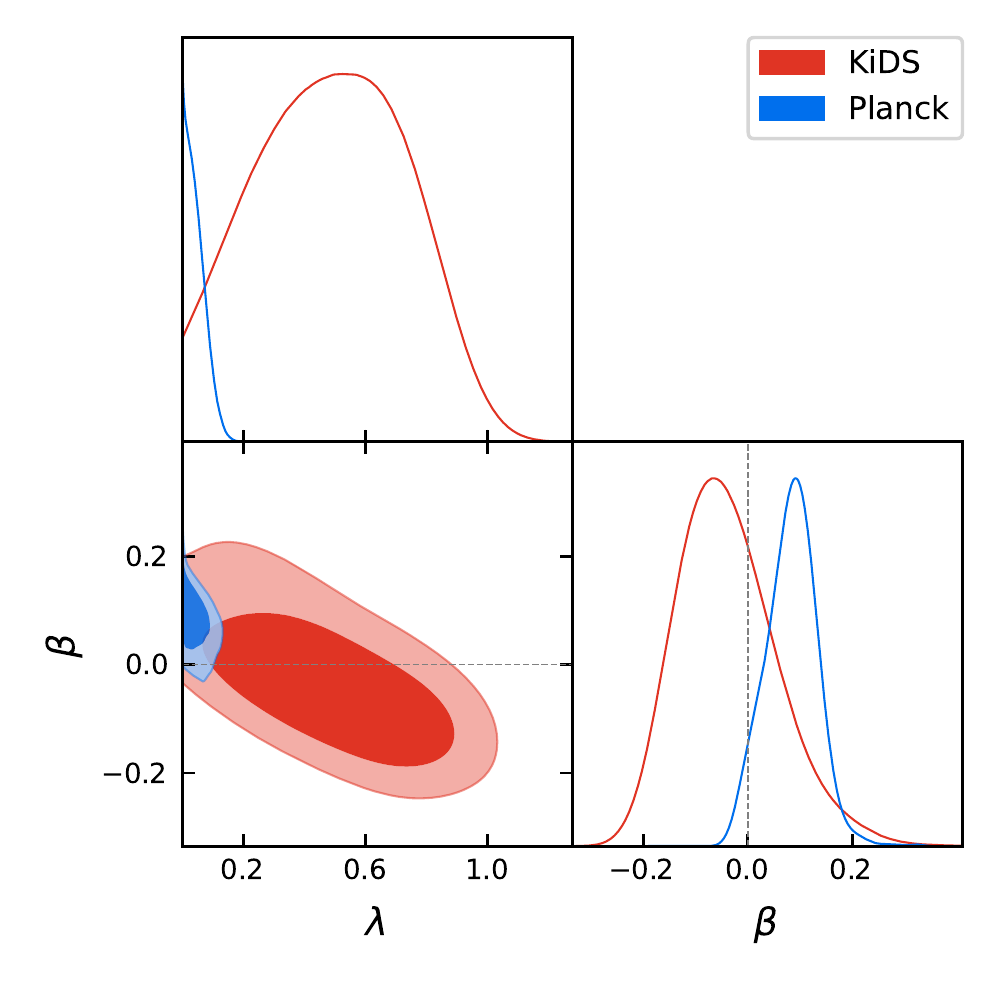}
\end{center}
\caption{Constraints on the ($\lambda$,$\beta$) posterior plane. Contours with 68$\%$ and 95$\%$  probability.}
\label{fig:specific}
\end{figure}

The probability contours and posterior distribution we obtain in the parameter space are displayed on Fig.~\ref{fig:specific} for the two specific degrees of freedom of the model. On the one hand, the parameter estimation done with Planck provides an upper bound for the posterior $\lambda$ which is compatible with a cosmological constant, that cannot be discarded.  The coupling posterior corresponds to the energy injection from dark energy to dark matter. However, on the other hand, weak lensing data brings a different constraint on the posterior $\lambda$, not favouring a vanishing value. It also slightly detects energy transfer from dark matter to dark energy.

We also undertake the same likelihood analysis with the standard model as a benchmark to compare the goodness-of-fit of the parametrisation (Table \ref{tab:comparison}). In light of the reduced chi-square, the two models look more or less equivalent. Nonetheless, according to the Akaike Information Criterion (AIC) \cite{1100705}, since it further penalises model complexity, the favourite cosmology for both probes is $\Lambda$CDM, this is particularly true for KiDS.

\begin{table}
\tbl{Standard $\Lambda$CDM model vs scalar field parametrisation}
{\begin{tabular}{@{}ccccc@{}}
\toprule
& \multicolumn{2}{c}{Planck}  & \multicolumn{2}{c}{KiDS}\\
  & \multicolumn{1}{c}{$\Lambda$CDM}  & \multicolumn{1}{c}{$\phi$ param.}   & \multicolumn{1}{c}{$\Lambda$CDM}  & \multicolumn{1}{c}{$\phi$ param.}\\
\colrule
$\chi^2$ & $602.6$ & $599.7$ & $321.5$ & $321.7$ \\ 
$\chi^2_{\rm{red}}$ & $1.0$ & $1.0$ & $2.6$ & $2.7$ \\
$AIC$ & $616.6$ & $617.7$ & $335.5$ & $339.7$ \\
$\Delta AIC$ & $0$ & $1.1$ &  $0$ & $4.2$ \\
\botrule
\end{tabular}
}
\label{tab:comparison}
\end{table}

\section{Concluding remarks}
\label{sec:pertubations}

Despite its simplicity, the parametrisation we propose is able to cover a large span of dark energy evolution with a limited number of parameters. Constraints on more complex parametrisations are usually limited by the degeneracies between the various degrees of freedom. The model preserves the fact that dark matter and dark energy are in comparable ratios today, while relaxing the initial conditions thanks to the existence of tracking solutions. It is thus capable of reproducing the evolution of the background, as well as the CMB anisotropies and the formation of Large Scale Structures. Since the parametrisation has a $\Lambda$CDM limit for small $\lambda$ and vanishing coupling $\beta$, it allows a direct test of departures from the standard model. We therefore argue that the parametrisation does help constrain the evolution of the equation of state and the departure from $\Lambda$CDM, even though the approach is purely phenomenological. Finally, the MCMC analysis carried out at perturbation level identifies tensions on the posteriors between early and late time probes. Background data is not expected to provide stronger constraints but the upcoming Euclid data will certainly improve the situation.

\section*{Acknowledgements}
The authors thank Giuseppe Fanizza and Miguel Zumalacarregui for their useful comments. This research was supported by Fundação para a Ciência e a Tecnologia (FCT) through the research grants: UID/FIS/04434/2019, PTDC/FIS-OUT/29048/2017 (DarkRipple), COMPETE 2020:POCI-01-0145-FEDER-028987 \& FCT:PTDC/FIS-AST/28987/2017 (CosmoESPRESSO) and IF/00852/2015.

\bibliographystyle{ws-procs961x669}
\bibliography{vdf.bib}

\begin{thebibliography}{10}

\bibitem{acel1}
A.~G. Riess {\em et~al.}, {Observational evidence from supernovae for an
  accelerating universe and a cosmological constant}, {\em Astron. J.} {\bf
  116}, 1009  (1998).

\bibitem{acel2}
S.~Perlmutter {\em et~al.}, {Measurements of Omega and Lambda from 42 high
  redshift supernovae}, {\em Astrophys. J.} {\bf 517}, 565  (1999).

\bibitem{amendola}
L.~Amendola and S.~Tsujikawa, {\em {Dark Energy}} (Cambridge University Press,
  2015).

\bibitem{mod}
T.~Clifton, P.~G. Ferreira, A.~Padilla and C.~Skordis, {Modified Gravity and
  Cosmology}, {\em Phys. Rept.} {\bf 513}, 1  (2012).

\bibitem{cc1}
P.~J.~E. Peebles and B.~Ratra, {The Cosmological constant and dark energy},
  {\em Rev. Mod. Phys.} {\bf 75}, 559  (2003), [,592(2002)].

\bibitem{Planck2018}
{Planck Collaboration}, {Planck 2018 results. VI. Cosmological parameters},
  {\em arXiv e-prints} , p. arXiv:1807.06209 (Jul 2018).

\bibitem{martin}
J.~Martin, {Everything You Always Wanted To Know About The Cosmological
  Constant Problem (But Were Afraid To Ask)}, {\em Comptes Rendus Physique}
  {\bf 13}, 566  (2012).

\bibitem{scalar3}
P.~J.~E. Peebles and B.~Ratra, {Cosmology with a Time Variable Cosmological
  Constant}, {\em Astrophys. J.} {\bf 325}, p. L17  (1988).

\bibitem{param_z}
J.~Weller and A.~Albrecht, Future supernovae observations as a probe of dark
  energy, {\em Physical Review D} {\bf 65} (May 2002).

\bibitem{param_a1}
M.~Chevallier and D.~Polarski, {Accelerating Universes with Scaling Dark
  Matter}, {\em {International Journal of Modern Physics D}} {\bf 10}, 213
  (2001), (uses Latex, 12 pages, 6 Figures) Minor corrections, Figures 4, 6
  revised. Conclusions unchanged.

\bibitem{param_a2}
E.~Linder, Exploring the expansion history of the universe, {\em Physical
  review letters} {\bf 90}, p. 091301 (04 2003).

\bibitem{Corasaniti}
P.~S. Corasaniti and E.~J. Copeland, Model independent approach to the dark
  energy equation of state, {\em Physical Review D} {\bf 67} (Mar 2003).

\bibitem{Wetterich:1994bg}
C.~Wetterich, {The Cosmon model for an asymptotically vanishing time dependent
  cosmological 'constant'}, {\em Astron. Astrophys.} {\bf 301}, 321  (1995).

\bibitem{Caldwell_1998}
R.~R. Caldwell, R.~Dave and P.~J. Steinhardt, Cosmological imprint of an energy
  component with general equation of state, {\em Physical Review Letters} {\bf
  80}, p. 1582–1585 (Feb 1998).

\bibitem{tsuq}
S.~Tsujikawa, {Quintessence: A Review}, {\em Class. Quant. Grav.} {\bf 30}, p.
  214003  (2013).

\bibitem{Nunes:2003ff}
N.~J. Nunes and J.~E. Lidsey, {Reconstructing the dark energy equation of state
  with varying alpha}, {\em Phys. Rev.} {\bf D69}, p. 123511  (2004).

\bibitem{Class1}
J.~Lesgourgues, The cosmic linear anisotropy solving system (class) i: Overview
   (2011).

\bibitem{Class2}
D.~Blas, J.~Lesgourgues and T.~Tram, The cosmic linear anisotropy solving
  system (class). part ii: Approximation schemes, {\em Journal of Cosmology and
  Astroparticle Physics} {\bf 2011}, p. 034–034 (Jul 2011).

\bibitem{Barreiro:1999zs}
T.~Barreiro, E.~J. Copeland and N.~J. Nunes, {Quintessence arising from
  exponential potentials}, {\em Phys. Rev.} {\bf D61}, p. 127301  (2000).

\bibitem{PhysRevD.62.043511}
L.~Amendola, Coupled quintessence, {\em Phys. Rev. D} {\bf 62}, p. 043511
  (2000).

\bibitem{Ma_1995}
C.-P. Ma and E.~Bertschinger, Cosmological perturbation theory in the
  synchronous and conformal newtonian gauges, {\em The Astrophysical Journal}
  {\bf 455}, p.~7 (Dec 1995).

\bibitem{MP1}
B.~Audren, J.~Lesgourgues, K.~Benabed and S.~Prunet, {Conservative Constraints
  on Early Cosmology: an illustration of the Monte Python cosmological
  parameter inference code}, {\em JCAP} {\bf 1302}, p. 001  (2013).

\bibitem{MP2}
T.~Brinckmann and J.~Lesgourgues, {MontePython 3: boosted MCMC sampler and
  other features}  (2018).

\bibitem{Feroz_2008}
F.~Feroz and M.~P. Hobson, Multimodal nested sampling: an efficient and robust
  alternative to markov chain monte carlo methods for astronomical data
  analyses, {\em Monthly Notices of the Royal Astronomical Society} {\bf 384},
  p. 449–463 (Jan 2008).

\bibitem{Feroz_2009}
F.~Feroz, M.~P. Hobson and M.~Bridges, Multinest: an efficient and robust
  bayesian inference tool for cosmology and particle physics, {\em Monthly
  Notices of the Royal Astronomical Society} {\bf 398}, p. 1601–1614 (Oct
  2009).

\bibitem{Feroz_2019}
F.~Feroz, M.~P. Hobson, E.~Cameron and A.~N. Pettitt, Importance nested
  sampling and the multinest algorithm, {\em The Open Journal of Astrophysics}
  {\bf 2} (Nov 2019).

\bibitem{Buchner:2014nha}
J.~Buchner, A.~Georgakakis, K.~Nandra, L.~Hsu, C.~Rangel, M.~Brightman,
  A.~Merloni, M.~Salvato, J.~Donley and D.~Kocevski, {X-ray spectral modelling
  of the AGN obscuring region in the CDFS: Bayesian model selection and
  catalogue}, {\em Astron. Astrophys.} {\bf 564}, p. A125  (2014).

\bibitem{Lewis:2019xzd}
A.~Lewis, {GetDist: a Python package for analysing Monte Carlo samples}
  (2019).

\bibitem{collaboration2019planck}
P.~Collaboration, Planck 2018 results. v. cmb power spectra and likelihoods
  (2019).

\bibitem{Hildebrandt_2016}
H.~Hildebrandt, M.~Viola, C.~Heymans, S.~Joudaki, K.~Kuijken, C.~Blake,
  T.~Erben, B.~Joachimi, D.~Klaes, L.~Miller and et~al., Kids-450: cosmological
  parameter constraints from tomographic weak gravitational lensing, {\em
  Monthly Notices of the Royal Astronomical Society} {\bf 465}, p. 1454–1498
  (Nov 2016).

\bibitem{1100705}
H.~Akaike, A new look at the statistical model identification, {\em IEEE
  Transactions on Automatic Control} {\bf 19}, 716  (1974).

\end{thebibliography}

\end{document}